\documentstyle[aps,prl,epsfig,twocolumn]{revtex}
\begin{document}
\draft
\title{Entanglement and Collective Quantum Operations}
\author{Anthony Chefles}
\address{Department of Physical Sciences, University of Hertfordshire \\
       Hatfield AL10 9AB, Herts, UK \\ email: A.Chefles@herts.ac.uk}
\author{Claire R. Gilson}
\address{Department of Mathematics, University of Glasgow, Glasgow G12 8QQ, UK}
\author{Stephen M. Barnett}
\address{Department of Physics and Applied Physics, University of Strathclyde \\ Glasgow G4 0NG, UK}                                                                                                   \input epsf
\epsfverbosetrue
\maketitle

\begin{abstract}

We show how shared entanglement, together with classical communication and local quantum
operations, can be used to perform an arbitrary collective quantum operation upon $N$
spatially-separated qubits. A simple teleportation-based protocol for achieving this,
which requires $2(N-1)$ ebits of shared, bipartite entanglement and $4(N-1)$ classical
bits, is proposed. In terms of the total required entanglement, this protocol is shown to
be optimal for even $N$ in both the asymptotic limit and for `one-shot' applications.
 \\ *
\end{abstract}

Interactions between physical systems involve the transmission of information between
them.  The future state of any subsystem will depend not only upon its own history, but
also those of the other subsystems.  This influence naturally involves the transmission of
information.

In classical physics, this information is purely classical.  In quantum physics, it is
quantum information that is transmitted  between the subsystems.  Unlike classical
information, quantum information cannot be copied\cite{NoCloning}. This implies that any
quantum information transferred to some system must be lost by its source in the process.
If this transfer of information is incomplete, which it often is, the result is
entanglement between the systems.

Entanglement forms a crucial link between classical and quantum information.  Nowhere is
this made more explicit than in the transmission of quantum information by
teleportation\cite{Teleport}. As is well-known, this can only be achieved by sending
classical information and making use of entanglement shared by the sending and receiving
locations.

Interactions between quantum subsystems are represented as collective operations on the
state space of the entire system.  In this Letter, we show how shared entanglement (SE),
together with classical communication (CC) and local quantum operations (LQ), can be used
to perform an arbitrary collective operation upon $N$ spatially-separated $2$-level
quantum systems (qubits), using a simple teleportation-based protocol. This requires
$2(N-1)$ ebits of bipartite entanglement to be shared between the locations of the qubits.

Large amounts of entanglement are difficult to produce under controlled circumstances, so
it is natural to enquire as to whether or not this figure is optimal.  For even $N$, we
give a graph-theoretic proof of the optimality of the teleportation protocol.  This holds
both for `one-shot' applications, where the operation is carried out only once, and also
in the asymptotic limit\cite{Concentration}, where the operation is carried out a large
number of times and we are interested in the average entanglement required per run of the
operation.

We begin  by considering the following scenario: take a network of $N$ laboratories,
$A_{j}$, where $j=1,{\ldots},N$, each of which contains a qubit.  We label these $q_{j}$.
The laboratories also share a certain amount of pure, bipartite entanglement with each
other. We shall refer to this as the {\em resource entanglement}.   Each $A_{j}$ also
contains auxiliary quantum systems, allowing arbitrary local collective operations to be
carried out in each laboratory. The laboratories can also send classical information to
each other.

Let us define the resource entanglement matrix ${\mathbf E}_{R}=\{E_{R}^{ij}\}$, where
$E_{R}^{ij}$ is the number of ebits shared by $A_{i}$ and $A_{j}$. This matrix is clearly
symmetric, has non-negative real elements and zeros on the diagonal.

From this matrix, we can construct a graph, which we term the {\em resource entanglement
graph}, $G_{E}(V,E)$. The vertex set $V$ is that of the laboratories $A_{j}$, and the edge
set $E$ represents the bipartite entanglement shared among them. The edge joining vertices
$A_{i}$ and $A_{j}$ has weight $E_{R}^{ij}$.  This weight is equal to the amount of pure,
bipartite entanglement shared by $A_{i}$ and $A_{j}$.  An edge of weight zero, which
represents no entanglement, is equivalent to no edge. The total resource entanglement is
\begin{equation}
E_{R}=\frac{1}{2}\sum_{ij}E_{R}^{ij}.
\end{equation}

We wish to use these resources to carry out an arbitrary collective operation upon the
$q_{j}$.   Perhaps the most natural way doing so is by teleportation. Teleportation of a
qubit from one location to another costs 1 ebit of entanglement and requires 2 classical
bits to be sent from the origin to the destination of the qubit\cite{Teleport}.

We can consider the situation in which all laboratories share entanglement and have the
resources for two-way classical communication with one particular laboratory.  Let this
laboratory be $A_{1}$.  The other laboratories can teleport the states of their qubits to
$A_{1}$.  The operation can then be carried out locally at $A_{1}$.  The final states of
the other qubits can then be teleported back to their original laboratories, completing
the operation.

This teleportation procedure requires each of the laboratories $A_{2},{\ldots},A_{N}$ to
share 2 ebits of entanglement with $A_{1}$ and for 2 bits of classical information to be
communicated each way between each of them and $A_{1}$.  The elements of the corresponding
resource entanglement matrix are
\begin{equation}
E_{R}^{ij}=2|{\delta}_{i1}-{\delta}_{1j}|.
\end{equation}
The corresponding graph $G_{E}$ is depicted in figure (1). The total resource entanglement
is
\begin{equation}
E_{R}=2(N-1).
\end{equation}
\begin{figure}[tbp]

\epsfxsize6cm \centerline{\epsfbox{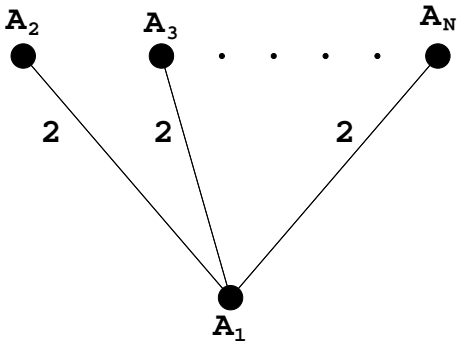}} \vspace{1cm} \caption{Resource entanglement
graph for the teleportation protocol.}

\end{figure}

Any quantum operation upon $N$ qubits can be performed using this method and thus, at
least for the topology of entanglement in our protocol, the value of $E_{R}$ in Eq. (3) is
sufficient.

This teleportation-based method for carrying out an arbitrary collective quantum operation
upon $N$ spatially separated qubits requires $E_{R}=2(N-1)$ ebits of entanglement.  Is
this figure optimal, in the sense that no less bipartite entanglement will suffice?

We can pose this question in the following, alternative way:  a network of laboratories
$A_{i}$ possesses shared bipartite entanglement, described by the graph $G_{E}$.  If the
corresponding total resource entanglement is sufficient to enable an arbitrary collective
operation to be performed, then what lower bound must $E_{R}$ satisfy?

The first observation we shall make is that if an arbitrary operation can be carried out
using the entanglement described by $G_{E}$, then any graph obtained from $G_{E}$ by a
permutation of the vertices also describes sufficient entanglement to carry out an
arbitrary operation. The permutation invariance of this sufficiency condition is
intuitive. We will provide a proof of it elsewhere \cite{Big}.

Consider the graph ${\tilde G}_{E}$ defined by
\begin{equation}
{\tilde G}_{E}=\sum_{P[V]}G_{E}(V).
\end{equation}
This graph is obtained from $G_{E}$ by summing over all permutations $P$ of the vertex set
$V$. By summing, we mean summing the entanglement represented by the weights of the edges.
The resource entanglement matrix ${\tilde {\mathbf E}}_{R}=\{{\tilde E}^{ij}_{R}\}$ for
this graph is easily obtained. Its elements are
\begin{equation}
{\tilde E}^{ij}_{R}=\sum_{P[V]}E_{R}^{P(i),P(j)}.
\end{equation}
This graph is regular and complete.  These properties follow immediately from the fact
that ${\tilde G}_{E}$, being defined as a sum over all vertex permutations, is itself
permutation invariant.

The total resource entanglement for this graph, ${\tilde E}_{R}$, is easily evaluated in
terms of the total resource entanglement of $G_{E}$.  There are $N!$ permutations of the
vertex set, implying that ${\tilde G}_{E}$ describes $N!$ times as much entanglement as
$G_{E}$, that is
\begin{equation}
{\tilde E}_{R}=N!E_{R}.
\end{equation}

All $N(N-1)/2$ edges in this graph have the same weight. Denoting this weight simply by
$e$, we obtain
\begin{equation}
e=2(N-2)!E_{R}.
\end{equation}

There are $N!$ permutations of the vertex set.  The permutation invariance of the
sufficiency condition then implies that the entanglement resources represented by ${\tilde
G}_{E}$ can be used to perform any operation $N!$ times.  By this, we mean the following:
suppose that $A_{i}$ contains $N!$ qubits.  We can then define $N!$ sets of qubits, where
each contains one from each laboratory.  It will be possible to perform the same operation
separately upon each of these sets.

Using the formalism we have set up, we can obtain the minimum value of $E_{R}$ exactly
when $N$ is even.  Our approach makes use of the SWAP operation upon 2 qubits. Consider a
pair of qubits, ${\alpha}$ and ${\beta}$, with respective states
$|{\psi}_{1}{\rangle}_{\alpha}$ and $|{\psi}_{2}{\rangle}_{\beta}$.  The SWAP operation,
$U_{S}$, exchanges the states of these subsystems:
\begin{equation}
U_{S}|{\psi}_{1}{\rangle}_{\alpha}{\otimes}|{\psi}_{2}{\rangle}_{\beta}=|{\psi}_{2}{\rangle}_{\alpha}{\otimes}|{\psi}_{1}{\rangle}_{\beta}.
\end{equation}
The property of $U_{S}$ that is of particular interest to us is its ability to create 2
ebits of entanglement.  To see how, suppose that in the laboratory containing ${\alpha}$
there is another qubit, ${\alpha}'$, and that these two qubits are initially prepared in a
maximally entangled state.  Likewise, ${\beta}$ is initially maximally entangled with a
neighbouring qubit ${\beta}'$.  If the SWAP operation is performed on ${\alpha}$ and
${\beta}$, then ${\alpha}$ will become maximally entangled with ${\beta}'$, and likewise
${\beta}$ and ${\alpha}'$ will become maximally entangled. Two ebits of entanglement have
been produced.

 The network of $N$ laboratories is assumed to possess
sufficient entanglement resources, described by the graph ${\tilde G}_{E}$, to enable any
operation to be carried out $N!$ times.  Here, we consider one particular operation, which
we will refer to as the pairwise-SWAP (PS) operation. Performing this operation once has
the effect of swapping the state of a qubit at $A_{j}$ with that of one at $A_{j+1}$, for
all odd $j$.  If we write the two-qubit SWAP operation exchanging the states of qubits
$q_{j}$ and $q_{j+1}$ as $U_{S}^{j+1,j}$, then the PS operation may be written as
\begin{equation}
U_{PS}=U_{S}^{N,N-1}{\otimes}U_{S}^{N-2,N-3}{\otimes}{\ldots}{\otimes}U_{S}^{2,1}.
\end{equation}
This operation is depicted in figure (2).

\begin{figure}[bp]

\epsfxsize5cm \centerline{\epsfbox{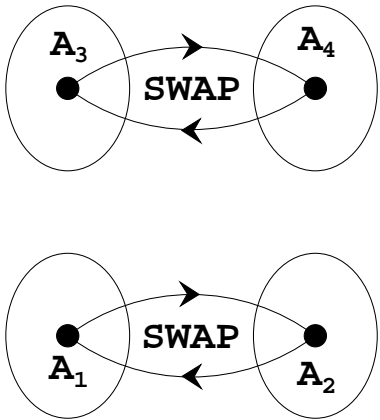}} \vspace*{1cm}

\caption{Depiction of the pairwise-SWAP (PS) operation for $N=4$.}

\end{figure}

The PS operation can then be used to  establish $N$ ebits of entanglement.  A fuller
discussion of multiqubit operations with this property will be given in \cite{Big}.  The
$N!$-fold PS operation can thus produce $N!N$ ebits of entanglement. Our aim is to use the
resources contained in the graph ${\tilde G}_{E}$ to perform the PS operation $N!$ times.
We wish to find the minimum value of $e$, and using Eq. (7), that of $E_{R}$, required to
do so.

To determine the minimum value of $e$ required to establish $2N!$ ebits of entanglement
between each pair of laboratories whose qubits' states will be exchanged by the PS
operation, we will make use of the fact that entanglement cannot increase under LQCC
operations.  Consider the situation depicted in figure (3).  We partition the entire
network into two sets. One contains the even laboratories $A_{2},A_{4},{\ldots},A_{N}$,
and the other contains the odd ones $A_{1},A_{3},{\ldots},A_{N-1}$.  We shall refer to
these sets as $S_{even}$ and $S_{odd}$.

\begin{figure}[tbp]
\epsfxsize6cm \centerline{\epsfbox{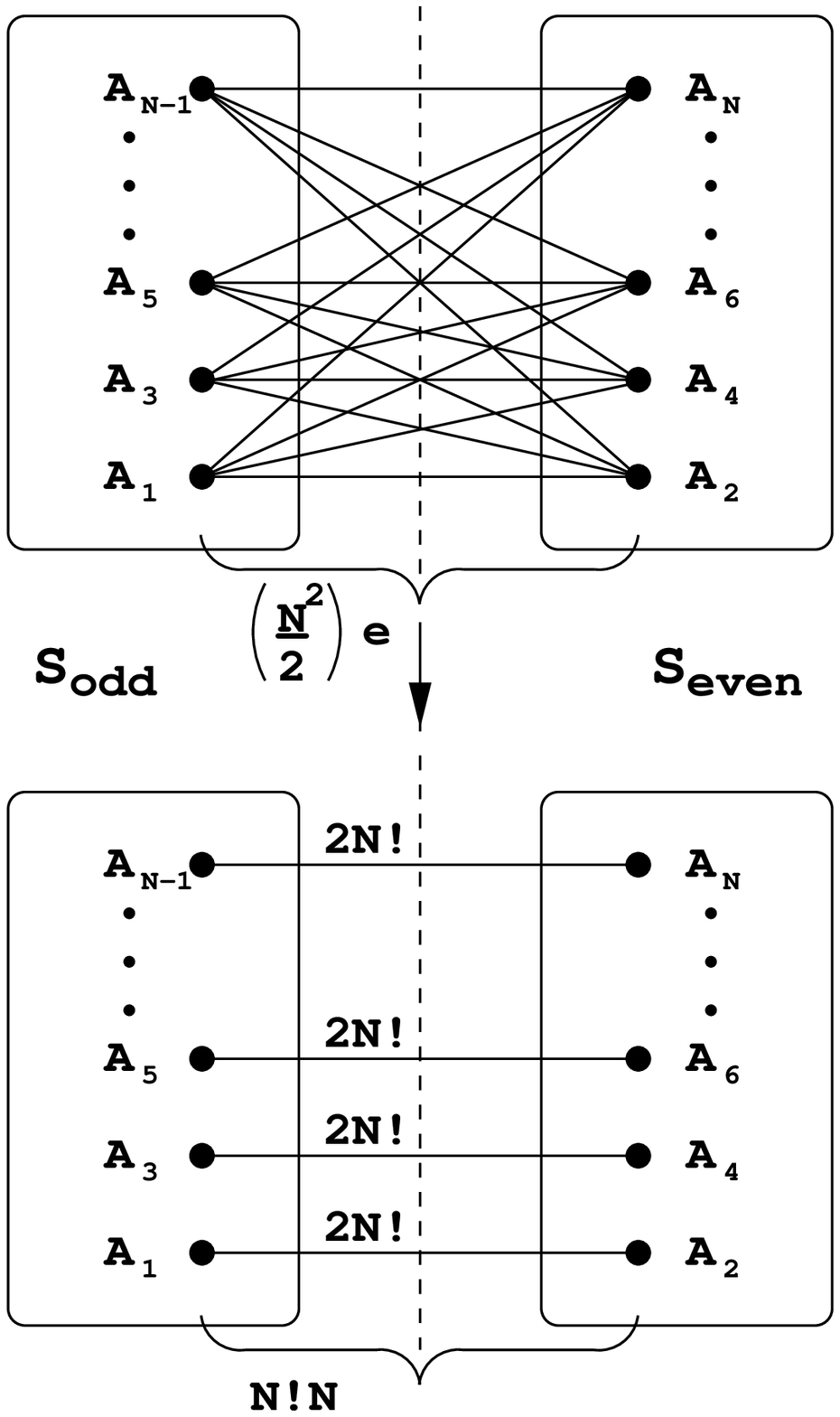}} \vspace*{1cm} \caption{Use of the resource
entanglement graph ${\tilde G}_{E}$ to carry out the $N!$-fold pairwise-SWAP operation.
Initially, the entanglement resources are distributed according to the graph ${\tilde
G}_{E}$. We have divided the $N$ laboratories into even and odd sets $S_{even}$ and
$S_{odd}$.  For the sake of clarity, we have not indicated the internal entanglement of
these sets. Each laboratory in $S_{odd}$ shares $e$ ebits of entanglement with each
laboratory in $S_{even}$.  These sets are separated by an imaginary partition, indicated
by the broken line.  Initially, these sets share $(N/2)^{2}e$ ebits, and the $N!$-fold PS
operation can create $N!N$ ebits.  The total entanglement shared across this partition
cannot increase, and the requirement that $e$ must be large enough to carry out the
$N!$-fold PS operation leads to inequalities (10) and (11). }

\end{figure}

According to the graph ${\tilde G}_{E}$, the total entanglement initially shared by these
sets can be calculated in a straightforward manner. Each of the $N/2$ laboratories in
$S_{odd}$ shares {\em e} ebits with each laboratory in $S_{even}$, that is, $Ne/2$ ebits
with $S_{even}$ in total. Adding up the $N/2$ such contributions from the laboratories in
$S_{odd}$ gives $(N/2)^{2}e$ ebits initially shared by $S_{even}$ and $S_{odd}$. The final
entanglement they share is $N!N$ ebits. The total entanglement that $S_{even}$ and
$S_{odd}$ share cannot increase, giving the inequality
\begin{equation}
\left(\frac{N}{2}\right)^{2}e{\geq}N!N.
\end{equation}
Making use of Eq. (7), we find that
\begin{equation}
E_{R}{\geq}2(N-1).
\end{equation}
This lower bound on the total resource entanglement is a tight bound, since this amount of
entanglement is precisely that which is required by the teleportation protocol.  Thus, for
even $N$, the teleportation protocol is optimal with regard to the required total resource
entanglement.  Using the above approach one can also calculated a lower bound on the
resource entanglement for an odd number of qubits\cite{Big}

We have derived this bound solely on the basis of the fact that, in a multiparticle
system, the entanglement shared by two exhaustive subsets, which will be of bipartite
form, cannot increase under LQCC operations.

Although the entanglement initially shared by each pair of laboratories is in pure,
bipartite form, the transformation shown in figure (3) may, at some point, manipulate the
resource entanglement into, possibly mixed, multiparticle entanglement. This does not
affect our argument. If the final entanglement is in multiparticle form, then in order to
carry out the $N!$-fold PS operation, $A_{j}$ and $A_{j+1}$ will have to be able to {\em
distill} $2N!$ ebits of pure, bipartite entanglement.   The {\em total} distillable
entanglement between $S_{even}$ and $S_{odd}$ cannot increase, which leads to inequality
(10) and thus the teleportation bound in (11).

The nonincreasing of entanglement under LQCC operations is an asymptotic result.  It
follows that the teleportation protocol is asymptotically optimal for even $N$.  By
asymptotic\cite{Concentration}, we mean that, given a very large number of sets of
separated qubits, where the same, arbitrary operation is to be carried out on each set,
the teleportation protocol uses the minimum {\em average} entanglement that is required
per run of the operation.

In practical situations, it is often the resources required to carry out an operation
successfully just once that will be of interest.  For general information processing
tasks, the resources required in the `one-shot' scenario are at least equal to the
resources required asymptotically.  For the problem we have considered here, when $N$ is
even, the entanglement resources required in both scenarios are equal.  This is because
the teleportation protocol, which requires $2(N-1)$ ebits, can be used to carry out any
collective operation on $N$ qubits once.

The classical communication resources required to carry out an arbitrary collective
operation on $N$ qubits can be analysed using the same technique.  We will present a
detailed discussion of this matter in\cite{Big}, but take the opportunity to mention that
the teleportation protocol, which, in addition to $2(N-1)$ ebits of entanglement, also
requires $4(N-1)$ classical bits, is also optimal in terms of classical communication
resources for even $N$.

In this Letter we have determined the minimum amount of bipartite entanglement required to
carry out an arbitrary operation upon an even number of qubits.  It is natural to attempt
to solve the same problem for the odd case.  Unfortunately, we have not been able to find
a collective operation on an odd number of qubits which yields the minimum resource
entanglement in the same manner as the PS operation.  Many aspects of the problem for odd
$N$ will be discussed in \cite{Big}, including the determination of lower bounds on the
minimum entanglement and communication resources.  We also examine the consequences of the
assumption that it costs an ebit to move an ebit.  This leads to the optimality of the
teleportation protocol in the one-shot case for the resource entanglement except possibly
when $N=3$.

\section*{Acknowledgements.}

We would like to thank Sandu Popescu, Noah Linden, Osamu Hirota and Masahide Sasaki for
interesting discussions.  Part of this work was carried out at the Japanese Ministry of
Posts and Telecommunications Communications Research Laboratory, Tokyo, and we would like
to thank Masayuki Izutsu for his hospitality.  This work was funded by the UK Engineering
and Physical Sciences Research Council, and by the British Council.

\end{document}